\documentclass[usenatbib,usegraphicx,useAMS]{mn2e}
\usepackage{aasmacros}

\def\lsim{\mathrel{\hbox{\rlap{\hbox{\lower4pt\hbox{$\sim$}}}\hbox{$<$}}}}
\def\gsim{\mathrel{\hbox{\rlap{\hbox{\lower4pt\hbox{$\sim$}}}\hbox{$>$}}}}
\newcommand{\lcdm}[0]{$\Lambda$CDM~}
\newcommand{\fig}[1]{Fig.~\ref{#1}}
\newcommand{\tab}[1]{Table~\ref{#1}}
\newcommand{\sect}[1]{Section~\ref{#1}}
\newcommand{\eqn}[1]{Eqn.~\ref{#1}}
\newcommand{\gadgetthree}[0]{{\sc gadget-3}}
\newcommand{\galform}[0]{{\sc galform}~}
\newcommand{\subfind}[0]{{\sc subfind}}
\newcommand{\magv}[0]{$M_{\rm{V}}$}
\newcommand{\halomass}[0]{$M_{crit,200}=1.42\times10^{12} M_{\odot}$}
\newcommand{\gammaEoS}[0]{\gamma_{\rm eff}}
\newcommand{\tsn}[0]{$t_{8 M_{\odot}}$}

\title[The Baryons in the MW Satellites]{The Baryons in the Milky Way Satellites}

\author[O.~H.~Parry et al.]{O.~H.~Parry$^{1}$\thanks{E-mail:o.h.parry@durham.ac.uk},
V.~R.~Eke$^{1}$, C.~S.~Frenk$^{1}$ and T.~Okamoto$^{1,2}$\\
$^{1}$ Institute for Computational Cosmology, Department of Physics,
University of Durham, Science Laboratories, South Road, Durham DH1
3LE\\
$^{2}$ Center for Computational Sciences, University of Tsukuba, 1-1-1 
 Tennodai, Tsukuba 305-8577 Ibaraki, Japan} 

\begin{document}

\date{Accepted 201? . Received 201? ; in original form 201?}

\pagerange{\pageref{firstpage}--\pageref{lastpage}}
\pubyear{201?}

\maketitle
\label{firstpage}

\begin{abstract}
We investigate the formation and evolution of satellite galaxies using
smoothed particle hydrodynamics (SPH) simulations of a Milky Way
(MW)-like system, focussing on the best resolved examples, analogous
to the classical MW satellites. Comparing with a pure dark matter
simulation, we find that the condensation of baryons has had a
relatively minor effect on the structure of the satellites' dark
matter halos.  The stellar mass that forms in each satellite agrees
relatively well over three levels of resolution (a factor of $\sim64$
in particle mass) and scales with (sub)halo mass in a similar way in
an independent semi-analytical model.  Our model provides a relatively
good match to the average luminosity function of the MW and M31.  To
establish whether the potential wells of our satellites are realistic,
we measure their masses within observationally determined half-light
radii, finding that they have somewhat higher mass-to-light ratios
than those derived for the MW dSphs from stellar kinematic data;
the most massive examples are most discrepant.  A statistical test yields
a $\sim6$ percent probability that the simulated and observationally
derived distributions of masses are consistent.  If the satellite
population of the MW is typical, our results could imply that feedback
processes not properly captured by our simulations have reduced the
central densities of subhalos, or that they initially formed with lower  
concentrations, as would be the case, for example, if the dark matter
were made of warm, rather than cold particles.
\end{abstract}


%%%%%%%%%%%%%%%%%%%%%%%%%%%%%%%%%%%%%%%%%%%%%%%%%%%%%%%%%%%%%%%%%%%%%%%%%%%%%%%%
%%%%%%%%%%%%%%%%%%%%%%%%%%%%%%%%%%%%%%%%%%%%%%%%%%%%%%%%%%%%%%%%%%%%%%%%%%%%%%%%
\section{Introduction} \label{sec:intro}
Substantial progress has been made over the last few years in
modelling the formation of galactic dark matter halos using high resolution
N-body simulations \citep{Springel2008,Diemand2008,Stadel2009}.
Hydrodynamical simulations of such systems inevitably lag behind in
terms of resolution, but are now reaching a point where they can be
used to investigate the detailed evolution of the baryonic component
of satellite galaxies, as demonstrated by several recent studies.
\citet{Okamoto2010a} studied the effects of different feedback models
on the chemical properties and luminosities of the satellite
populations around three Milky Way (MW)-mass halos.
\citet{OkamotoFrenk2009} showed that a combination of the early
reionisation of pregalactic gas at high redshift and the injection of
supernovae energy is sufficient to suppress star formation in the
myriad of low mass subhalos that form in the \lcdm cosmology,
confirming results from earlier semi-analytic modelling
\citep{Benson2002,Somerville2002}.  \citet{Wadepuhl2010} further
argued that cosmic rays generated by supernovae may play an important
role in suppressing star formation in satellites.

From an observational point of view, the release of data from the
Sloan Digital Sky Survey (SDSS) \citep{York2000} over the last decade
has transformed the study of the Local Group satellites.  The $\sim30$ faint dwarf
galaxies discovered using those data
\citep[e.g.,][]{Zucker2004,Martin2006,Belokurov2007,McConnachie2008} 
have prompted a new phase of detailed testing of current galaxy formation
theories on smaller scales and in more detail than ever before.  The
SDSS data also reduced the discrepancy that existed between the
number of low mass dark matter halos predicted by the \lcdm
cosmological model and the number of faint satellites identified 
around the MW: the `missing satellite problem'
\citep{Klypin1999,Moore1999}. Over the same period, numerous
theoretical models
\citep[e.g.,][]{LiDeLuciaHelmi2010,Maccio2010,Font2011} have confirmed
early conclusions that a combination of a photoionising background and
feedback processes from supernovae (SNe) are capable of bringing the
two into good agreement. 

However, it is important to recognise that the satellite problem is
not simply a statement that star formation must be suppressed in low
mass halos.  A more subtle, but equally important test of any
cosmological model is whether the potential wells in which satellite
galaxies form are capable of supporting stellar systems with realistic
kinematics.  \citet{StrigariFrenkWhite2010} demonstrated that all of
the classical MW satellites for which high quality kinematic data are
available are consistent with having formed in dark matter subhalos
selected from the high resolution \lcdm N-body simulations of the
Aquarius project \citep{Springel2008}.  However, successful models
must satisfy both constraints, producing realistic luminosity
functions \emph{and} forming stars in potentials like those inferred
from observations. 

\citet{Boylan-Kolchin2011} have recently argued that the most massive 
subhalos in high resolution simulations of cold dark matter halos are
too concentrated to be able to host the brightest observed satellites
of the Milky Way.  \citet{Lovell2011} have shown that subhalos formed
from warm, rather than cold, dark matter have suitably low
concentrations, but both they and Boylan-Kolchin et al. emphasise that
other solutions to the discrepancy are possible.  One promising
possibility is the mechanism originally proposed by Navarro, Eke \&
Frenk (1996a), whereby the rapid expulsion of dense central gas in a
starburst can unbind the inner parts of the halo, significantly 
reducing its concentration. We find an example of this process in one
of the subhalos formed in the simulations analysed in this paper.

From a theoretical perspective, {\it ab initio} hydrodynamic
simulations are uniquely well-suited to investigating the
effects of galaxy formation on the dark matter halos of satellite galaxies.
It has been known for some time that baryons may significantly alter the behaviour
of dark matter on some scales.  Dissipative processes such as gas
cooling, star formation and feedback decouple the dynamical evolution
of the baryons from that of the dark matter.  The resulting change in
the shape of the overall potential in turn affects the phase 
space structure of the dark matter halo.  

Central concentrations of cold baryonic matter can induce an
adiabatic, radial contraction of the central regions
\citep{Blumenthal1986,Gnedin2004}, while the opposite effect can be
achieved if dense clumps of baryonic material heat the central
distribution of dark matter
\citep{MoMao2004,Mashchenko2006,Mashchenko2008}, or if the blowout
mechanism of Navarro, Eke and Frenk is effective
\citep[see
also][]{GelatoSommerlarsen1999,GnedinZhao2002,MoMao2004,ReadGilmore2005,Governato2010,PontzenGovernato2011}.
This latter mechanism was originally proposed as a means to erase the
central dark matter cusps in dwarf galaxies, though whether or not
such `cored' profiles are required by the observations remains a
matter of ongoing debate
\citep{Goerdt2006,SanchezSalcedo2006,Strigari2006,Gilmore2007,Walker2009}. 
If baryons really do modify the dark matter in satellites on sub-kpc
scales significantly, then the value of studying dwarf galaxies with
post-processed N-body simulations may be very limited.

In this paper, we make use of a model that has already had success in
reproducing some properties of the Local Group satellites, including the
shape and approximate normalisation of their luminosity function and
the metallicity-luminosity relation \citep{Okamoto2010a}.  In
\sect{sec:SIM} we outline the details of our simulations, including the initial conditions,
simulation code and modelling of various key baryonic physical
processes.  In \sect{sec:DMO}, we expand on the theoretical predictions
of the model by examining what effect baryons have had on the dark
matter profiles of satellites.  We perform tests in
\sect{sec:CONV} to ensure that key properties of our satellite
population do not depend on the numerical resolution.  In \sect{sec:obs} we compare several observable and derived
properties of the simulated satellites to Local Group data.  Finally,
in \sect{sec:nef_sub} we discuss the evolution of one particularly
interesting satellite in the simulation, which is dominated by its
stellar component at $z=0$.  Our main
results are summarised in \sect{sec:CONCLUSIONS}. 

%%%%%%%%%%%%%%%%%%%%%%%%%%%%%%%%%%%%%%%%%%%%%%%%%%%%%%%%%%%%%%%%%%%%%%%%%%%%%%%%
%%%%%%%%%%%%%%%%%%%%%%%%%%%%%%%%%%%%%%%%%%%%%%%%%%%%%%%%%%%%%%%%%%%%%%%%%%%%%%%%
\section{The simulations} \label{sec:SIM}
To investigate the properties of a simulated MW-satellite system, we
select one of the six halos from the Aquarius project described in
\citet{Springel2008}, halo `C' in their labelling system.   These
halos were extracted from a cosmological simulation in a cube of
comoving volume $\rm (100Mpc)^{3}$ and were chosen to have masses close to
that of the Milky Way ($\sim10^{12}M_{\odot}$) and avoid dense environments (no
neighbour exceeding half its mass within $1h^{-1}$Mpc)
\citep{Navarro2010}.

As in Aquarius, we employ a `zoom' resimulation technique, with higher
mass boundary particles used to model the large scale potential and
lower mass particles in a $\sim5 h^{-1}$Mpc region surrounding the
target halo.  Extra power is added to the initial particle
distribution on small scales in the high resolution region, as described
by \citet{Frenk1996}.  We assume a
\lcdm cosmology, with parameters $\Omega_{m}=0.25$,
$\Omega_{\Lambda}=0.75$, $\Omega_{b}=0.045$, $\sigma_{8}=0.9$,
$n_{s}=1$ and $H_{0}=100h {\rm kms}^{-1}{\rm Mpc}^{-1}=73{\rm
  kms}^{-1}{\rm Mpc}^{-1}$.

\begin{table}
  \centering
  \caption{Numerical parameters adopted for the three different
    resolution simulations: dark matter and gas particle
    masses and the gravitational softening in physical units.}
  \label{tab:SIM_PROPS}
  \begin{tabular}{cccc}
    &$\rm M_{DM}[M_{\odot}]$& $\rm M_{gas}[M_{\odot}]$&$\rm \epsilon_{phys}[pc]$\\
    \hline
    \hline
    Aq-C-4&$2.6\times10^5$&$5.8\times10^4$&257\\
    Aq-C-5&$2.1\times10^6$&$4.7\times10^5$&514\\
    Aq-C-6&$1.7\times10^7$&$3.7\times10^6$&1028\\
    \hline
  \end{tabular}
\end{table}

The highest resolution realisation of halo C in Aquarius had a dark
matter particle mass of $\rm 1.4\times10^4M_{\odot}$.  However, the
extra computational time associated with hydrodynamic simulations
makes such a resolution impractical; our highest resolution instead
corresponds to a dark matter particle mass of $\rm
\sim2.6\times10^5M_{\odot}$ and an initial gas particle mass of $\rm
5.8\times10^4M_{\odot}$.  In order to conduct convergence studies, we
also simulated the halo at two lower resolutions, with particle masses
$\sim8$ and $\sim64$ times larger.  We adopt the same naming
convention as \citet{Springel2008}, labelling the three runs (in order
of decreasing resolution) Aq-C-4, Aq-C-5 and
Aq-C-6. \tab{tab:SIM_PROPS} lists the numerical parameters of each
simulation.

Our simulation code is based on an early version of the PM-Tree-SPH
code \gadgetthree.  Baryonic processes are modelled as described in
\citet{Okamoto2010a}, with a number of modifications designed to improve
the treatment of supernovae-driven winds.  In the following
subsections, we summarise some of the most important features of the
code with emphasis on aspects that have the greatest impact on
satellite formation.

%%%%%%%%%%%%%%%%%%%%%%%%%%%%%%%%%%%%%%%%%%%%%%%%%%%%%%%%%%%%%%%%%%%%%%%%%%%%%%%%
\subsection{Radiative cooling and the equation of state}\label{sec:SIM_COOLING}

Radiative processes in our model are implemented as described in
\citet{WiersmaSchayeSmith2009} and include inverse Compton scattering of CMB
photons, thermal Bremsstrahlung, atomic line cooling and
photoionisation heating from Hydrogen and Helium.  All gas in the
simulation volume is ionised and heated by a spatially uniform, time
evolving UV background, as calculated by \citet{HaardtMadau2001}.
During the reionisation of H and He I ($z=9$) and He II ($z=3.5$), an
extra two eV per atom of thermal energy is added to the gas, smoothed
over Gaussian distributions with widths $\Delta z = 0.0001$ and $0.5$
respectively, in order to approximately account for non-equilibrium
and radiative-transfer effects
\citep{AbelHaehnelt1999}.  The contributions to heating and cooling
from eleven elements (H, He, C, N, O, Ne, Mg, Si, S, Ca and Fe) are
interpolated from tables output by {\sc cloudy} \citep{Ferland1998},
using elemental abundances smoothed over the SPH
kernel, to approximate the mixing of metals in the interstellar
medium (ISM).  This avoids unphysical small scale fluctuations in the
cooling time that arise if the abundances associated with individual
particles are used \citep{Wiersma2009}.

Failure to resolve the Jeans mass ($M_{J}$) or Jeans Length
($\lambda_{J}$) is known to lead to spurious fragmentation through
gravitational instability
\citep{BateBurkert1997,Truelove1997}.  Techniques employed to avoid
this problem typically involve some form of energy injection to
maintain an effective pressure that guarantees that the available
resolution is sufficient
\citep[e.g.,][]{Machachek2001,RobertsonKravtsov2008,SpringelHernquist2003,Ceverino2010,Schaye2010}.
This pressure support has been identified with, for example, a hot
phase maintained through energy input by SNe
\citep{SpringelHernquist2003} and turbulence induced by the disk's self
gravity and rotation \citep{WadaNorman2007}.  We include a minimum
pressure explicitly by adopting a polytropic equation of state (EoS) with
$P_{\rm{min}}\propto\rho^{\gammaEoS}$ for gas above the density threshold
for star formation and adopt $\gammaEoS=1.4$.  For $\gammaEoS > 4/3$, the Jeans
mass increases with density 
\citep[e.g.,][]{SchayeDVecchia2008}, such that, if it is resolved at
the threshold density, it is resolved everywhere.  \citet{Schaye2010}
showed that, in models where star formation is strongly regulated by
stellar feedback, as it is in ours, the choice of $\gammaEoS$ has very
little impact on the global star formation rate.  The star formation
threshold density is $n_{\rm H} >0.1{\rm cm}^{-3}$ for the Aq-C-6
simulation and a factor of four and sixteen higher for the Aq-C-5 and
Aq-C-4 simulations respectively.  This scaling is chosen since, for
irradiated primordial gas with an isothermal density profile, halving
the gravitational softening will increase the maximum density that is
resolved by a factor of four.

%%%%%%%%%%%%%%%%%%%%%%%%%%%%%%%%%%%%%%%%%%%%%%%%%%%%%%%%%%%%%%%%%%%%%%%%%%%%%%%%
\subsection{Multiphase ISM and Star Formation}\label{sec:SIM_ISM_SFR}

At sufficiently high pressures, the ISM is known to exist in distinct
phases.  We follow \citet{SpringelHernquist2003}, modelling
gas above the threshold density using hybrid SPH particles, which are assumed to
consist of a series of cold clouds in pressure equilibrium with a
surrounding hot phase.  The total mass in the cold phase can increase through
thermal instability and decrease through star formation and cloud
evaporation by SNe, but the mass spectrum of clouds is kept fixed as  
\begin{equation}
  \Phi(m)= \frac{dN_{c}}{dm}\propto m^{-\alpha}.
  \label{eqn:cloudmassfunc} 
\end{equation}
We adopt $\alpha=1.7$, guided by observations that suggest a
plausible range of $1.5 - 1.9$
\citep{SolomonRivolo1989,Fukui2001,HeyerCarpenterSnell2001}. 
We follow \citet{SamlandGerhard2003} in assuming clouds to be
spherical, with size at a fixed mass determined solely by the
ambient pressure \citep{Elmegreen1989}:
\begin{equation}
  \left(\frac{m}{\rm M_{\odot}}\right)\left(\frac{r(m)}{\rm pc}\right)^{-2}= 190P_{4}^{1/2},
  \label{eqn:cloudsizes} 
\end{equation}
where $P_{4} = \frac{P/k}{10^{4}{\rm K cm^{-3}}}$ and  k is the Boltzmann
constant. The dependence of each cloud's dynamical time on the
effective pressure follows directly from \eqn{eqn:cloudsizes}:
\begin{equation}
  t_{dyn} = \left[\frac{3\pi}{32G\rho(m)}\right]^{1/2} \simeq 0.32
  P_{4}^{-3/8}\left(\frac{m}{M_{\odot}}\right)^{1/4} {\rm Myr},
  \label{eqn:cloudtdyn} 
\end{equation}
and we assume that the star formation rate in each cloud is inversely proportional
to its dynamical time ($t_{dyn}$):
\begin{equation}
  \dot{m}_{*}= c_{*}\frac{m}{t_{dyn}},
  \label{eqn:cloudSFR} 
\end{equation}
where $c_{*}$ is the star formation efficiency, which is set to reproduce
the normalisation of the Kennicutt-Schmidt law \citep{Kennicutt1998}.
The total star formation rate for each SPH particle is obtained by
integrating Eqn.~\ref{eqn:cloudSFR} over all clouds deemed
capable of supporting star formation (which we assume to be in the mass
range $10^{4}-10^{6}M_{\odot}$).  

SPH particles spawn new collisionless star particles in a stochastic fashion,
with a probability that depends on their star formation rate and on
the mass in the cold phase.  Each star particle represents a single
stellar population, forming with a \citet{Chabrier2003} initial mass
function (IMF).  Energy, mass and metals are returned to the ISM by
AGB stars, type Ia and type II SNe on timescales appropriate for the
age and metallicity of the stellar population, with yields and stellar
lifetimes taken from \citet{Portinari1998} and \citet{Marigo2001}.
Rather than performing this calculation at every dynamical timestep,
which is very expensive computationally, we use coarser steps, chosen
such that the timescales associated with type II and type Ia SNe can
be adequately sampled.  Initially, each step has length \tsn$/50$,
where \tsn is the lifetime of an $\rm 8M_{\odot}$
star \footnote{\tsn$\sim40$ Myr at solar metallicity}.  When the age
of the stellar population exceeds \tsn, the timesteps lengthen to
$100{\rm Myr}$, which is short enough to model the release of mass and
energy from type Ia SNe and from AGB stars.

%%%%%%%%%%%%%%%%%%%%%%%%%%%%%%%%%%%%%%%%%%%%%%%%%%%%%%%%%%%%%%%%%%%%%%%%%%%%%%%%
\subsection{Supernovae Winds} \label{sec:SIM_WINDS}

The perennial problem with the distribution of stellar feedback energy
in cosmological hydrodynamical simulations has been that the
star-forming gas that receives the energy is dense enough to radiate
it away before it can have any dynamical effect
\citep{KatzWeinbergHernquist1996}. This is likely to
be a consequence 
of the inability to resolve the detailed structure of the ISM
\citep[e.g.,][]{DVecchiaSchaye2008,CeverinoKlypin2009}.
We employ a commonly used technique to circumvent this limitation,
which consists of imparting kinetic energy to gas particles directly
\citep[e.g.,][]{NavarroWhite1993,MihosHernquist1994,SpringelHernquist2003}.
The velocity we choose to give gas particles that receive SNe energy
is motivated by observations that suggest that large scale outflows have 
velocities that scale with the circular velocity of their host
galaxies \citep{Martin2005}.  As a proxy for the host halo's circular
velocity, which is computationally expensive to calculate for each
particle on-the-fly, we use the local one-dimensional velocity
dispersion, determined from neighbouring dark matter particles.  This
quantity is strongly correlated with the maximum circular velocity,
$v_{max}$, in a way that does not evolve with redshift
\citep{Okamoto2010a}.  Our prescription results in a wind speed that
increases as the halo grows and hence, from energy conservation, in a
mass loading (wind mass per unit star formation rate) that is highest
at early times. This scaling has been shown to give a much better
match to the luminosity function of the Milky Way satellites than
models that use a constant wind velocity \citep{Okamoto2010a}. 

One further addition to the model is needed to ensure that SNe driven
winds act as intended. \citet{DVecchiaSchaye2008} showed that standard
kinetic feedback is more effective in low mass galaxies, where wind
particles tend to drag neighbouring gas out with them.  In high mass
galaxies on the other hand, the pressure of the ISM can be sufficient
to prevent much of the mass in the wind from escaping.  Since we wish
to be able to prescribe the mass loading and wind velocity directly,
we choose to decouple wind particles from the hydrodynamic calculation
for a short time in order to allow them to escape the high density
star forming regions.  When the density has fallen to
$n_{H}=0.01cm^{-3}$, the particles feel the usual hydrodynamic force
again.  If they do not reach sufficiently low densities after a time
$\rm 10kpc/v_{wind}$, they are recoupled anyway.

When a gas particle receives SNe energy from a neighbouring
star particle, the wind speed ($v_{w}$) is obtained from the local
velocity dispersion and then the particle is assigned a probability to
be added to the wind: 
\begin{equation}
  p_{w}= \frac{\Delta Q}{ \frac{1}{2} m_{sph}v_{w}^{2}},
  \label{eqn:windprob} 
\end{equation}
where $\Delta Q$ is the total feedback energy received by the gas
particle and $m_{sph}$ is the current mass of the SPH particle.  Note
that an SPH particle's mass may increase if it receives mass from SNe
or AGB stars in neighbouring star particles, or decrease if it spawns
a new star particle, which has a mass of half the original gas
particle mass.  If $p_{w}$ exceeds unity, that is, if there is energy
available in excess of that needed to add the particle to the wind, then
the extra energy is distributed to the gas particle's neighbours as an
increase in internal energy.  The direction in which wind particles
are propelled is chosen at random to be parallel or anti-parallel to
the vector $(\vec{v_{0}}-\vec{\overline{v}})\times
\vec{a}_{grav}$ where $v_{0}$ is the velocity of the gas particle
before it receives feedback energy, $\vec{a}_{grav}$ is the
gravitational acceleration vector, pointing approximately to the local potential
minimum (halo centre) and $\overline{v}$ is the bulk velocity of the
halo, which we take to be the mean velocity of the gas particle's dark
matter neighbours. The result of this treatment is a wind 
launched preferentially along an object's rotation axis
\citep{SpringelHernquist2003}.  Our model for SNe winds differs from that
described by \citet{Okamoto2010a} in two ways.  Firstly, we allow all
gas particles, not just those above the star formation density
threshold, to be added to the wind if they receive feedback energy.  The original
prescription can result in a variable wind mass loading depending on
how well the star forming region is resolved.  Secondly, only type II
SNe contribute to the winds, type Ia SNe energy is added to the gas as thermal
energy.

%%%%%%%%%%%%%%%%%%%%%%%%%%%%%%%%%%%%%%%%%%%%%%%%%%%%%%%%%%%%%%%%%%%%%%%%%%%%%%%%
\subsection{Satellite Identification}
Galaxies are identified using a version of the \subfind\ algorithm
\citep{Springel2001} adapted by \citet{Dolag2009}, which identifies
self-bound structures and includes the internal energy of gas
when computing particle binding energies.  From the $\rm \sim5Mpc$ high
resolution region, we select all galaxies within 280kpc of the centre
of the most massive (central) galaxy.   This distance was chosen to
match the limiting magnitude of the completeness-corrected satellite
luminosity function constructed by \citet{Koposov2008}.  The largest 
satellite in our Aq-C-4 run is resolved with about $1.5\times10^5$ particles in
total, $\sim3\times10^4$ of which are star particles. In the
following, we consider all galaxies with more than ten star particles,
which, taking into account the typical mass fraction lost through
stellar evolution for our choice of IMF, implies a stellar mass limit of $\rm
\sim1.2\times10^{5}M_{\odot}$ for Aq-C-4.  

%%%%%%%%%%%%%%%%%%%%%%%%%%%%%%%%%%%%%%%%%%%%%%%%%%%%%%%%%%%%%%%%%%%%%%%%%%%%%%%%
%%%%%%%%%%%%%%%%%%%%%%%%%%%%%%%%%%%%%%%%%%%%%%%%%%%%%%%%%%%%%%%%%%%%%%%%%%%%%%%%
\section{The Effect of Baryons on Satellite Dark Matter Halos} \label{sec:DMO}

Using a dark matter only (DMO) counterpart of our Aq-C-4 run,
simulated as part of the Aquarius project \citep{Springel2008}, we
have examined the extent to which the dynamics of the baryons alter
the structure of dark matter (sub)halos of satellite galaxies over the
course of their formation.  The DMO run had identical initial
conditions to our Aq-C-4, but for the absence of baryons and a
correspondingly higher dark matter particle mass by a factor $\sim
1/(1-\Omega_{b}/\Omega_{m})$.

Naively, one might simply compare each subhalo with its
DMO equivalent at $z=0$, but this turns out to be problematic.  As has
been noted in previous N-body simulations at different resolutions,
small phase deviations in subhalo orbits get amplified over time, such that subhalos can be in
quite different positions at $z=0$ ~\citep[e.g.,][]{Frenk1999,Springel2008}.  We
see similar differences between Aq-C-4 and the DMO run.  Subhalo orbits
are also affected by other factors such as subhalo-subhalo scattering
and variations in the potential due to small differences in the growth
history of the main halo.

Since the strength of tidal shocking is strongly dependent on
pericentric distance \citep{Gnedin1999,Mayer2001}, small orbital
deviations can cause large differences in subhalo structure, which are
entirely unrelated to the presence or absence of baryons.  This
complication can be avoided, either by choosing subhalos with no close
pericentre, or by making the comparison at the epoch when the
satellite is first accreted into the halo of the main galaxy, before
the orbits have had a chance to diverge. We choose the latter option,
since the former restricts us to a very small number of cases,
although we note that one massive halo in a low eccentricity
($\sim0.2$) orbit with a distant pericentre ($\rm \sim200kpc$) shows
comparatively small differences in its dark matter density profile at
$z=0$ relative to the DMO case.  In the few instances where the
accretion times of the subhalo differ slightly between the
hydrodynamical and DMO runs, we choose the earlier of the two epochs.

\begin{figure*}
  \includegraphics[width=0.8\textwidth,angle=270]{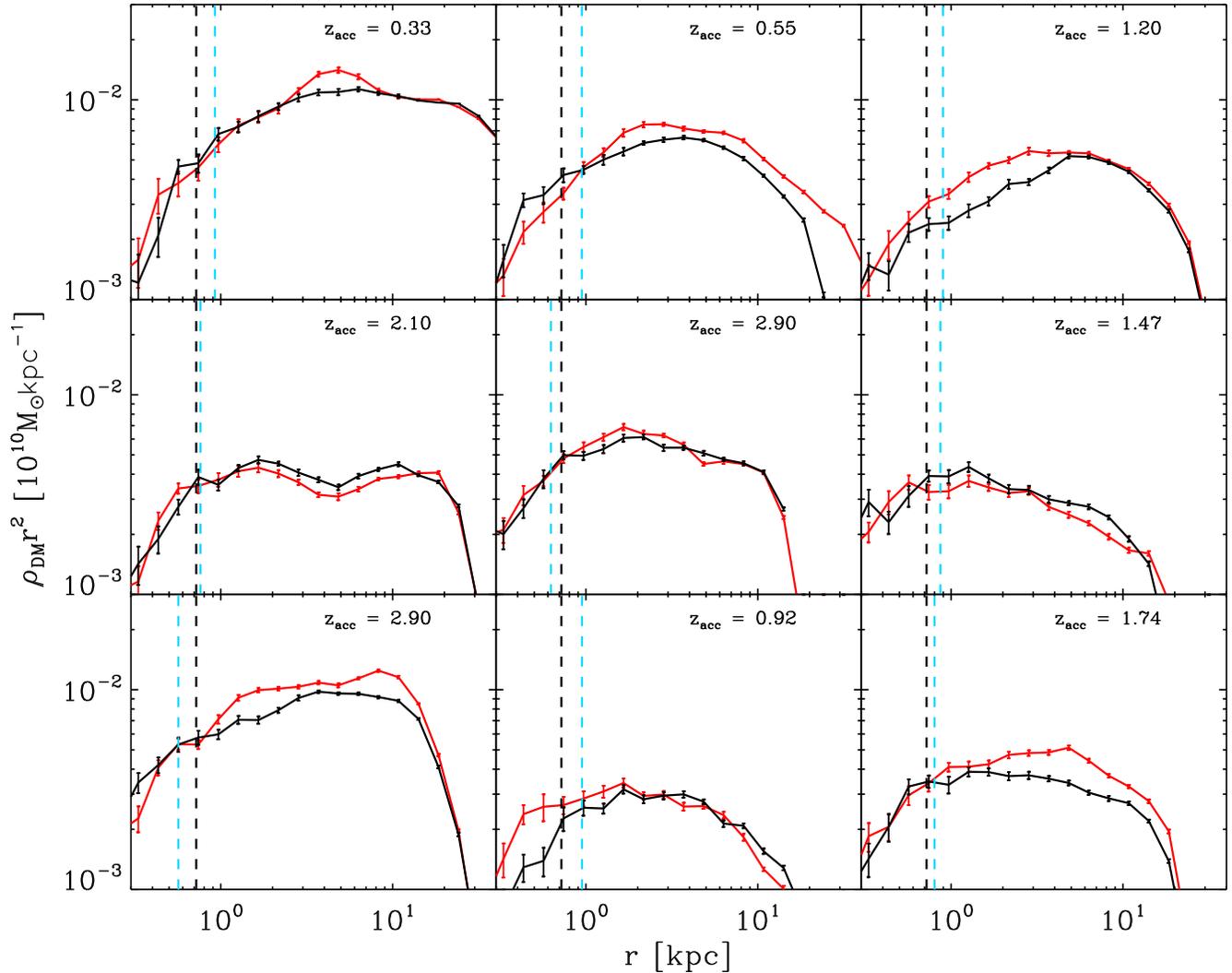}
\caption{The spherically averaged dark matter density
  profiles of the nine most massive satellite galaxies in our high
  resolution hydrodynamic (red) and dark matter only (black) runs.
  Error bars are obtained by bootstrap resampling.  The comparisons
  are made at the redshift where the galaxy is first accreted as a
  satellite, which is shown as a label at the top of each panel.
  Black dashed vertical lines indicate the scale on which softened
  gravitational forces become fully Newtonian.  Blue dashed vertical
  lines indicate the convergence radius of \citet{Power2003}.}
\label{fig:DMrhoprofs}
\end{figure*}

\begin{figure*}
  \includegraphics[width=0.8\textwidth,angle=270]{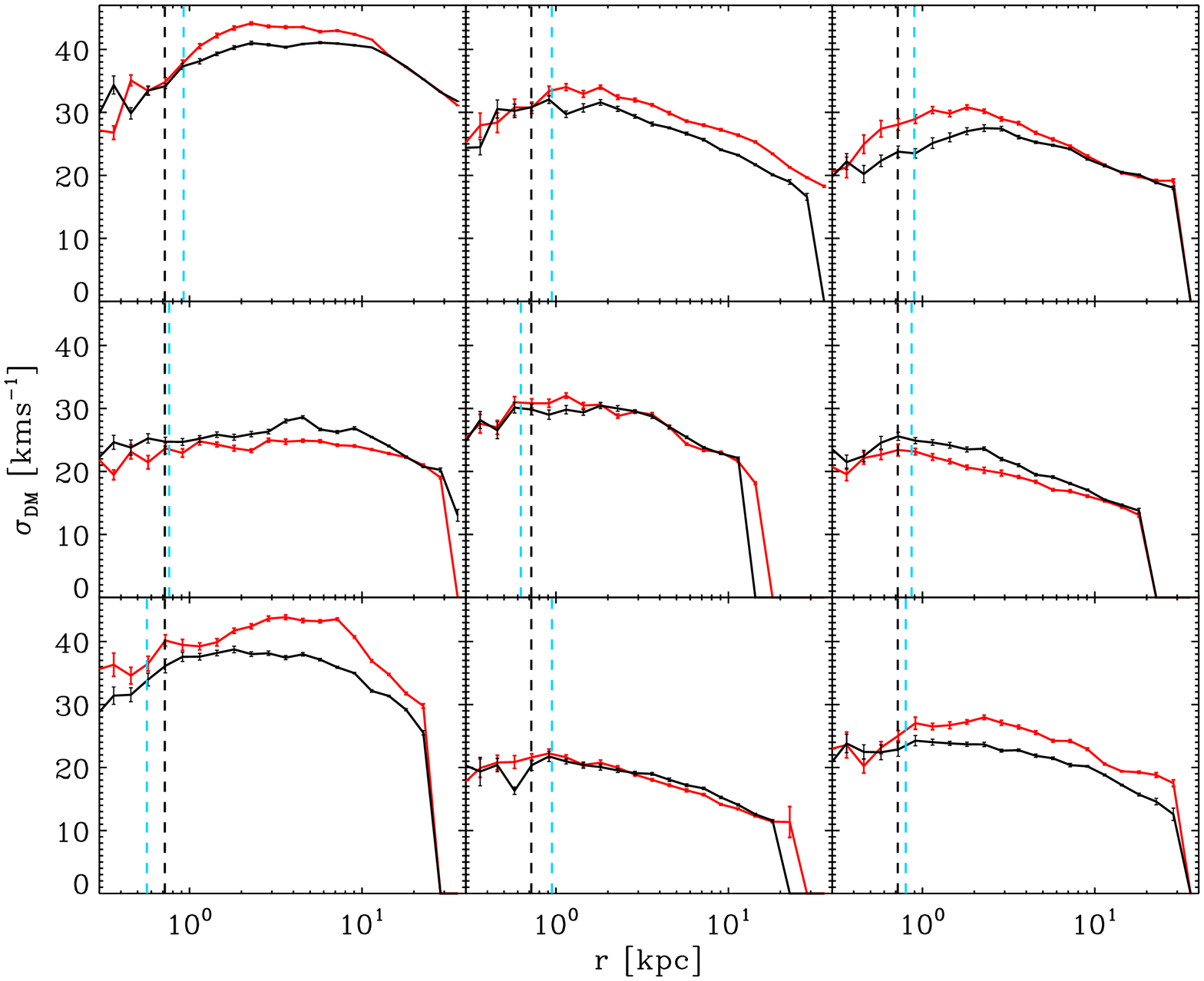}
\caption{The spherically averaged 1D velocity dispersion profiles
  of the most massive satellite galaxies in our high resolution
  hydrodynamic (red) and dark matter only (black) runs.  Error bars
  are obtained by bootstrap resampling.  The comparisons are made at
  the redshift when the galaxy is first accreted as a satellite, which
  is indicated on the corresponding panels in \fig{fig:DMrhoprofs}.
  Black dashed vertical lines indicate the scale on which softened
  gravitational forces become fully Newtonian.  Blue dashed vertical
  lines indicate the convergence radius of \citet{Power2003}.}
\label{fig:DMsigprofs}
\end{figure*} 

In \fig{fig:DMrhoprofs} we show spherically averaged profiles for the
dark matter density (plotted as $\rho r^{2}$ to emphasise small
differences) of our most massive satellites, for Aq-C-4 and the DMO run
at the output time when each satellite first joins the main friend-of-friends
\citep[FOF;][]{Davis1985} group.\footnote{In practice, we require the 
subhalo to be counted in the main FOF group for two consecutive
snapshots to avoid instances where subhalos are spuriously joined to
the main group for a short time.}  The differences in the subhalo
density profiles with and without baryons clearly exceed the
uncertainties associated with finite sampling, indicated by the error
bars.  They are also greater than, for example, the differences
expected between dark matter realisations of the same subhalo at
different resolutions \citep{Springel2008}.  Some subhalos (e.g. top
left and centre right panels) appear to have been largely unaffected,
whilst others (e.g. top and bottom right panels) show more substantial
changes of up to thirty percent in some radial bins.  In \sect{sec:nef_sub},
we describe an extreme example of a subhalo which suffered much more
extensive damage as a result of baryonic processes. In general,
however, there does not seem to be any consistent trend for baryons to
increase or decrease the central density of the dark matter.

In \fig{fig:DMsigprofs}, we show spherically averaged dark matter
velocity dispersion profiles for the same selection of satellites in
Aq-C-4 and the DMO run.  The largest differences are seen in those
subhalos that show the most change in their density profiles in
\fig{fig:DMrhoprofs}.  Once again, the differences are typically less 
than ten percent in any given radial bin, but as much as thirty
percent in some instances, with no apparent trend for baryons to raise
or lower the dark matter velocity dispersion.

From these results, we conclude that the baryons have had a relatively
small impact on the dark matter phase-space structure of the subhalos,
with the important caveat that it is unclear whether such effects are
limited by the resolution of our simulations.  Another important
factor contributing to this conclusion is the strength of our
feedback, since it dictates how easily baryons are able to condense in 
the centre of low mass halos and affect the dynamics of the dark
matter. Models with much weaker feedback might achieve more pronounced
differences than we see here, but as we will show in
\sect{sec:LF}, such models typically overpredict the luminosities of
satellite galaxies.

Although some satellites are accreted at fairly high redshift (see
the labels on each panel in \fig{fig:DMrhoprofs}), it is unlikely that
the baryons would have an increased effect in the remaining time to $z=0$. In all 
but the largest satellites, gas is lost fairly rapidly following
accretion, as we will show in \sect{sec:massevo}.

%%%%%%%%%%%%%%%%%%%%%%%%%%%%%%%%%%%%%%%%%%%%%%%%%%%%%%%%%%%%%%%%%%%%%%%%%%%%%%%%
%%%%%%%%%%%%%%%%%%%%%%%%%%%%%%%%%%%%%%%%%%%%%%%%%%%%%%%%%%%%%%%%%%%%%%%%%%%%%%%%
\section{Convergence of Satellite Properties} \label{sec:CONV}
In this section we investigate the convergence of various key
properties of our simulated satellite galaxies.  We make the
comparison on an object-to-object basis, matching up satellites
between runs.  As explained in \sect{sec:DMO}, this cannot be
accomplished simply by choosing subhalos that are spatially closest at
the final time.  Instead, we trace particles back to the initial
conditions and match them spatially there.  In the following analysis
we consider the most massive satellites in Aq-C-4 for which resolved
counterparts exist in Aq-C-5 and often also in Aq-C-6.

%%%%%%%%%%%%%%%%%%%%%%%%%%%%%%%%%%%%%%%%%%%%%%%%%%%%%%%%%%%%%%%%%%%%%%%%%%%%%%%%
\subsection{Stellar Mass} \label{sec:mstars}

We begin by considering the total stellar mass in each satellite at
$z=0$.  This is a function of the rate at which gas can cool onto the
galaxy and the efficiency of star formation, dictated by the gas
physics and feedback.  As well as checking that our results do not
depend on resolution, we compare with an independent modelling
technique, presented by \citet{Cooper2010}, hereafter C10. They used
the Aquarius simulations to track the formation of dark matter substructures in
six different halos and a version of the semi-analytic galaxy
formation code \galform to compute the baryonic properties of
satellite galaxies.\footnote{Their semi-analytic model is
essentially that presented by \citet{Bower2006}, but with a lower
circular velocity threshold ($\rm 30~kms^{-1}$) to identify halos in
which cooling is suppressed by reionisation.  This value is motivated
by recent hydrodynamical
simulations~\citep{Hoeft2006,OkamotoGaoTheuns2008}.}  At each output
time, the stellar mass formed since the last simulation snapshot is
assigned to some fraction of the most tightly gravitationally bound
dark matter particles in the subhalo, providing spatial and kinematic
information for the stars.  The `tagged' fraction was chosen to match
the distribution of sizes (half-light radii) for Local Group
satellites and also to produce results robust to changes in
resolution.

\begin{figure}
\begin{center}
\leavevmode
\includegraphics[width=0.475\textwidth]{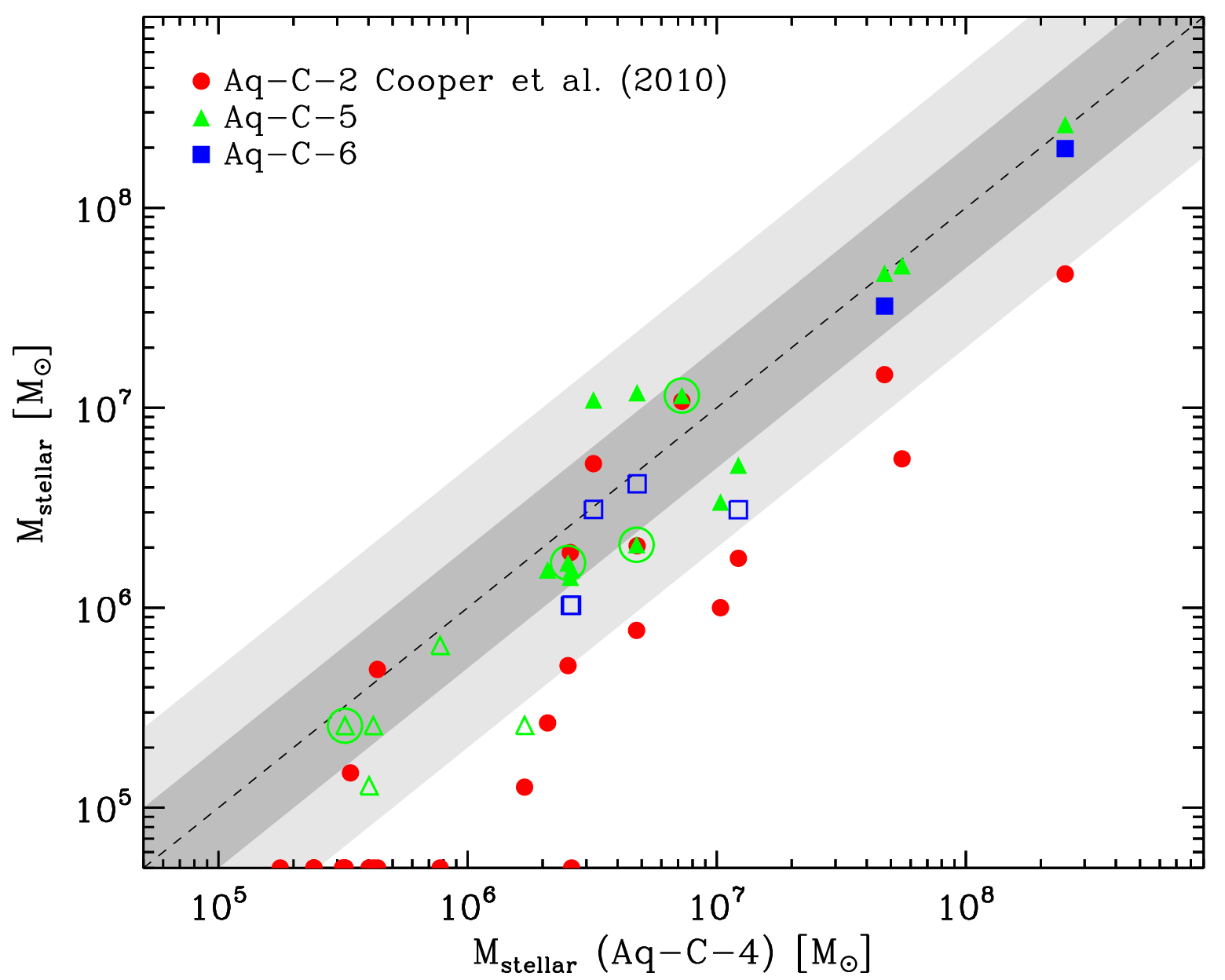}
\end{center}
\caption{A comparison of the stellar mass that forms in each satellite
  at the three different resolutions of our hydrodynamical runs, as
  well as in the independent semi-analytic model presented by
  \citet{Cooper2010}.  Note that not all satellites in the high
  resolution run have resolved counterparts at intermediate and low
  resolution. Points that lie on the abscissa correspond to subhalos
  with $\rm M_{stellar}<5\times10^{4}M_{\odot}$.  Data
  points are plotted for all satellites in the high resolution run
  with more than ten star particles.  If the satellite also has ten
  star particles in the lower resolution run, it is plotted as a
  filled, rather than open symbol.  Satellites that have lost more
  than 50 percent of their maximum stellar mass through tidal
  stripping are indicated by circled points.  The dark and light gray
  shaded regions represent factors of two and five respectively away
  from the line of equality.}
\label{fig:mstar}
\end{figure}

\fig{fig:mstar} demonstrates that the stellar mass in each subhalo
agrees relatively well between the three resolutions, particularly for
the most massive examples, although the difference is as large as a
factor of six in one case.  Some of this scatter (between different
resolutions and between the semi-analytic and hydrodynamical
realisations) is likely related to the deviations in subhalo orbits
between simulation runs described in \sect{sec:DMO}.  Small differences
in pericentric distance and eccentricity can strongly affect the tidal
field and hence the extent to which stars can be stripped from
subhalos.  Examples of subhalos that have lost more than 50 percent of
their peak stellar mass (excluding the effects of stellar evolution)
in one or more of the runs are indicated in \fig{fig:mstar} by circled
points.

The semi-analytic prescription typically predicts a lower stellar
mass in each subhalo, although the correlation indicates that the
ranking of subhalos by stellar mass is similar to that in our
simulations.  This offset between the two techniques is not obviously
attributable to a single aspect of either model, but one 
mechanism that may be important is ram pressure stripping.  In the
semi-analytic model, any hot gas is instantaneously stripped from
satellites upon infall.  In combination with strong SNe feedback, this
quenches star formation in satellites very rapidly following
accretion.  As we will show in the next subsection, ram pressure and
feedback act to produce a similar effect in our simulations, but star
formation is able to continue for an appreciable time after accretion
and right up to $z=0$ for the most massive satellites. 

%%%%%%%%%%%%%%%%%%%%%%%%%%%%%%%%%%%%%%%%%%%%%%%%%%%%%%%%%%%%%%%%%%%%%%%%%%%%%%%%
\subsection{Mass evolution} \label{sec:massevo}

\begin{figure*}
\begin{center}
\leavevmode
\includegraphics[width=0.8\textwidth,angle=270]{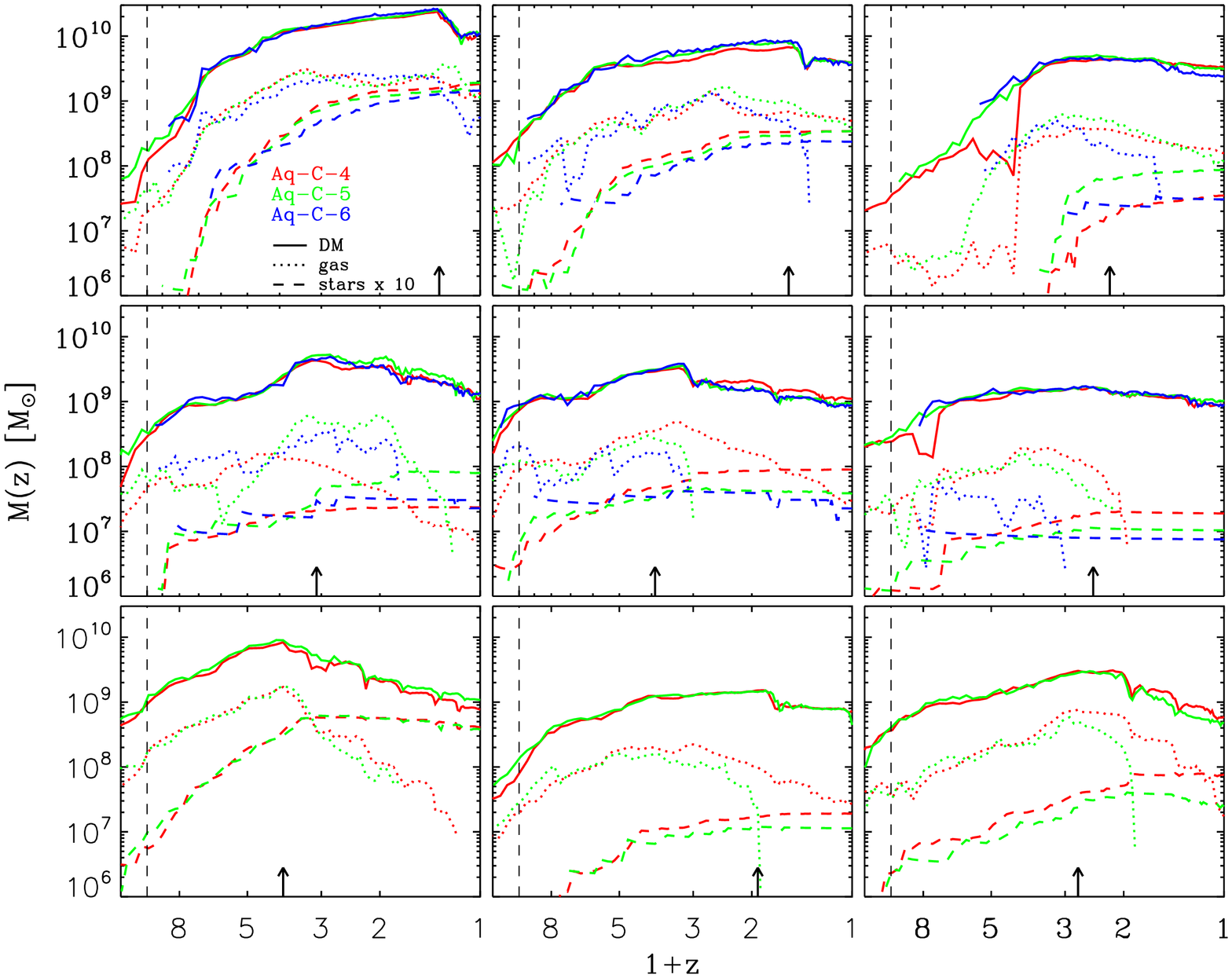}
\end{center}
\caption{The mass evolution in dark matter (solid lines), stars
  (dashed lines) and gas (dotted lines) for the nine most massive
  satellite galaxies in our Aq-C-4 (red), Aq-C-5 (green) and Aq-C-6
  (blue) simulations.  Note that no resolved counterpart was
  identified in the Aq-C-6 simulation for the satellites tracked in
  the bottom row of panels.  Stellar masses are scaled up by a factor
  of ten to reduce the range of the ordinate axis. The vertical dashed
  line indicates the redshift of reionisation and arrows the epoch at
  which the galaxy was first accreted as a satellite in Aq-C-4.  The
  panels are ordered by total mass, from left to right along each row,
  beginning at the top.}
\label{fig:massevo}
\end{figure*}

We now examine how satellites in the three runs acquire their dark and
baryonic mass and form stars. \fig{fig:massevo} shows the
gravitationally bound mass of dark matter, stars and gas for the nine
most massive satellites as a function of redshift.  For reference, we
also include \tab{tab:masses}, which lists the masses at $z=0$. 
Reionisation ($z=9$) is marked with a vertical dashed line and arrows
indicate the accretion time, that is, the time when the galaxy first
becomes a satellite in the high resolution run.  Where the accretion
times differ slightly between runs, the value for the high resolution
case is shown.

Apart from numerical convergence, which we will discuss next,
there are a number of interesting features to note in
\fig{fig:massevo}, many of which were also observed in the simulations
of \citet{Okamoto2010a}, \citet{Wadepuhl2010} and \citet{Sawala2011}.
There are instances of satellites being periodically stripped of mass
as they pass through pericentre, most obviously the satellite tracked
in the bottom right panel, which loses dark matter and gas from its
outer parts in two close approaches.  None of these massive satellites
appear to be on orbits with sufficiently high eccentricity and/or a
close pericentre to strip the more tightly bound stellar component,
although there are several examples of satellites in the simulation
that were heavily stripped or disrupted entirely and hence are not
among the most massive at $z=0$.

The loss of gas following accretion in most cases is fairly rapid and
is brought about through a combination of ram pressure stripping and
stellar feedback. As noted by both \citet{OkamotoFrenk2009} and
\citet{Wadepuhl2010}, reionisation appears to have virtually no
impact on satellites this large (the lowest mass example in
\fig{fig:massevo} has final total mass of $\rm
5\times10^{8}M_{\odot}$); the gas mass rises steadily through $z=9$,
and star formation continues unabated.  We note that the three largest
satellites retain a substantial amount of gas and are still increasing
their stellar mass at the present day, analogously to the ongoing star
formation in the MW dIrrs.

In terms of numerical convergence, in most respects, there is good
agreement between the three resolutions.  The exceptions to this are
the rate at which gas is lost from some satellites following accretion
and the resulting effect on the late-time star formation rates. In
most cases, there is a clear tendency for more efficient ram pressure
stripping with decreasing resolution.  This affect appears to be
related to the force resolution, which results in gas particles in
lower resolution runs being less tightly bound and hence more
susceptible to ram pressure stripping.  The slightly different
timescales over which satellites are able to retain their gas and
continue to form stars account, at least in part, for the often lower
final stellar masses in Aq-C-5 and Aq-C-6 noted in the previous
subsection.

\begin{table}
\centering
\begin{tabular}{ccc}
$M_{DM}$ & $M_{gas}$ & $M_{*}$\\
$\rm [10^{6}M_{\odot}]$ & $\rm [10^{6}M_{\odot}]$ & $\rm [10^{6}M_{\odot}]$\\
\hline
\hline
14822.3 & 1734.8 & 250.5\\
14236.2 & 584.1 & 27.7\\
5436.6 & 678.1 & 47.2\\
4634.8 & 215.2 & 4.8\\
1548.1 & 0.0 & 3.2\\
1496.5 & 40.6 & 12.2\\
1229.1 & 0.0 & 2.6\\
1049.7 & 0.0 & 55.4\\
1001.8 & 37.6 & 2.6\\
\hline
\end{tabular}
\caption{The mass in dark matter, gas and stars gravitationally bound to the nine most massive satellites at $z=0$}
\label{tab:masses}
\end{table}

%%%%%%%%%%%%%%%%%%%%%%%%%%%%%%%%%%%%%%%%%%%%%%%%%%%%%%%%%%%%%%%%%%%%%%%%%%%%%%%%
%%%%%%%%%%%%%%%%%%%%%%%%%%%%%%%%%%%%%%%%%%%%%%%%%%%%%%%%%%%%%%%%%%%%%%%%%%%%%%%%
\section{Observed Properties} \label{sec:obs}
In the previous section, we demonstrated that our model produces satellite
galaxies with properties that show reasonable convergence with resolution and
stellar masses that scale with subhalo mass in a fashion expected from
an alternative modelling technique.  We now proceed to examine how
well their observable properties match those of the Local Group
satellites.  Where photometric quantities are required, we use the stellar
population synthesis model PEGAS$\acute{E}$ \citep{Fioc1997}, summing
the luminosities of all star particles gravitationally bound to the
subhalo at $z=0$.

%%%%%%%%%%%%%%%%%%%%%%%%%%%%%%%%%%%%%%%%%%%%%%%%%%%%%%%%%%%%%%%%%%%%%%%%%%%%%%%%
\subsection{Satellite Luminosity Function} \label{sec:LF}
One of the most fundamental properties of any galaxy population is its
luminosity function. Encoded in its shape and normalisation are a
range of physical processes that are  key to understanding the
formation and evolution of the population. 

\begin{figure}
\begin{center}
\leavevmode
\includegraphics[width=0.475\textwidth]{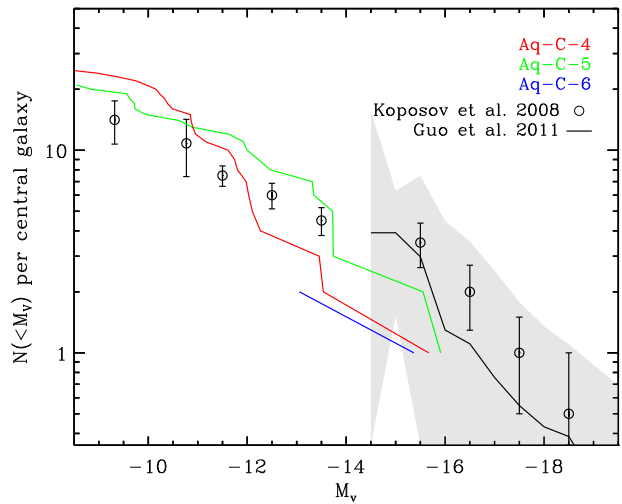}
\end{center}
\caption{The rest frame, V-band luminosity function of satellite
  galaxies in our high (red), intermediate (green) and low (blue)
  resolution simulations. Circles with error bars indicate the
  average MW+M31 satellite luminosity function, corrected for completeness, by
  \citet{Koposov2008}.  For consistency with their data, only
  simulated satellites within 280kpc of the central galaxy are
  included.  The solid black line and shaded regions indicate the
  mean and spread in the satellite luminosity functions of Milky Way
  mass galaxies in an analysis of SDSS data by \citet{Guo2011}.
  The light grey region shows the RMS scatter about the mean in
  each bin.}
\label{fig:LF}
\end{figure}

In \fig{fig:LF} we plot the luminosity function of simulated
satellites at each resolution.  Also plotted is the
(completeness-corrected) average luminosity function for MW and M31
satellites from \citet{Koposov2008} that includes the SDSS ultra-faint
objects.  We find good agreement between resolutions at all resolved
luminosities, consistent with the convergence of stellar masses and
star formation histories demonstrated in \sect{sec:CONV}.  The
simulated populations provide good matches to the Local Group average
at the faint end, consistent with the findings of \citet{Okamoto2010a}
in another of the Aquarius halos, but have no galaxy as bright as the
Large Magellanic Cloud (LMC).  The brightest galaxy (in Aq-C-4) has a
similar V-band magnitude to that of the Small Magellanic Cloud (SMC).

Note that there is still a significant observational uncertainty in
the total mass of the MW. Current estimates put it between
$0.8$ and $\rm 3\times10^{12} M_{\odot}$
\citep{Dehnen2006,LiWhite2008,Xue2008,Watkins2010}.  Our halo has a
mass slightly closer to the lower end of this range of
\halomass.  Clearly, if we are simulating a halo twice
or half as massive as the MW, we should not expect to reproduce the
luminosity function of its satellites exactly.  A further
consideration is whether the MW has a typical satellite population for
its halo mass or total luminosity.  To this end, we have also plotted
a black solid line and two shaded regions indicating the mean and
spread of the luminosity function (see figure caption for details) for
satellite systems in the SDSS around central galaxies with r-band
luminosity close to that of the MW
\citep{Guo2011}.  These data suggest that the Local Group is fairly
typical, although satellites like the Magellanic Clouds are found in
fewer than half of the systems in the sample.  In another study,
\citet{Liu2010} found that $>80$ percent of MW-like galaxies have no
satellite as bright as the SMC within 150 kpc.  In this statistical
context, perhaps the lack of very bright satellites in our simulations
should not be a major cause for concern.

%%%%%%%%%%%%%%%%%%%%%%%%%%%%%%%%%%%%%%%%%%%%%%%%%%%%%%%%%%%%%%%%%%%%%%%%%%%%%%%%
\subsection{Sizes} \label{sec:sizes}

An important and readily observable property of the most massive Local
Group satellites is the distribution of their sizes, usually measured
as the radius containing half the luminosity in projection.
Unfortunately, a combination of the spatial resolution and the
limitations of our subgrid model for star formation mean that we
cannot hope to reproduce the observed sizes in our simulations.  As
explained in \sect{sec:SIM}, a minimum pressure is maintained in star
forming gas to ensure that the Jeans length on the equation of state,
$\lambda_{J,EoS}$, is always resolved.  Our subgrid model assumes that
stars form on much smaller (unresolved) scales, inside molecular
clouds, but the star particles that are created must nonetheless
inherit the dynamical properties of the SPH particle from which they
formed.  As a result, the minimum size of star-forming regions will be
dictated by the warm/hot phase density and temperature, through
$\lambda_{J,EoS}$, or by the gravitational softening,  if this is
larger.

\begin{figure}
\begin{center}
\leavevmode
\includegraphics[width=0.4\textwidth]{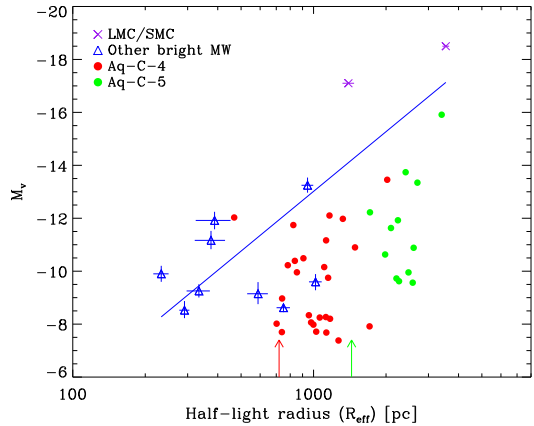}
\end{center}
\caption{Absolute V-band magnitude as a function of de-projected
    half-light radius for satellite galaxies in our Aq-C-4 (red filled
    circles) and Aq-C-5 (green filled circles) runs.  The Magellanic
    Clouds are represented by purple crosses.  Their projected
    half-light radii were derived from angular values taken
    from \citet{BothunThompson1988} and distance estimates from
    \citet{HilditchHowarthHarries2005} and \citet{Pietrzynski2009}.
    3D half-light radii were calculated by multiplying these values by 4/3, as
    suggested by \citet{Wolf2010}.  All other MW satellites brighter
    than $-7.5$ (except Saggitarius) are shown as blue triangles with
    luminosity and de-projected half-light radii taken from
    \citet{Wolf2010}.  The solid blue line is a least-squares fit to
    all the observational data points.  Arrows indicate the scales on
    which softened gravitational forces become fully Newtonian for
    each resolution.}
\label{fig:sizes}
\end{figure}

These limits are apparent in \fig{fig:sizes}, where we plot the
absolute V-band magnitude of simulated satellites as a function of
their half-light radius.  The observed half-light radii are
de-projected by multiplying by a factor of 4/3, an approximation that
is accurate to $2\%$ for the exponential, Gaussian, 
King, Plummer and Sersic profiles commonly used to fit the MW
satellites \citep{Wolf2010}.  The stars in both the Aq-C-4 and Aq-C-5
runs typically have much less concentrated distributions than the
observed satellites.  An exception to this is the third brightest
satellite in Aq-C-4, which has a V-band magnitude of $-12.2$ and a
half-light radius of $\sim480$pc.  It has a very high mass fraction in
stars and an unusual history, forming in a series of violent major
mergers at $z\sim4$ before being subjected to strong tidal disruption
between $z=2$ and $z=0$.  We discuss this satellite in detail in
\sect{sec:nef_sub}.

In the highest resolution run, Aq-C-4, we expect the gravitational
softening to be the main factor limiting the minimum sizes of star
forming regions, since it is always larger than $\lambda_{J,EoS}$.  In
Aq-C-5 and Aq-C-6 (not shown here), which have lower threshold
densities for star formation by factors of four and sixteen
respectively, $\lambda_{J,EoS}$ at the threshold is comparable to the
softening, so should also be important in setting the sizes of the
stellar component. For both Aq-C-4 and Aq-C-5, the half-light radius
of the most massive galaxy should not be limited by either
effect and is consistent with the observations, given the large
scatter.

%%%%%%%%%%%%%%%%%%%%%%%%%%%%%%%%%%%%%%%%%%%%%%%%%%%%%%%%%%%%%%%%%%%%%%%%%%%%%%%%
\subsection{Dynamical Masses}

\begin{figure}
\includegraphics[width=0.4\textwidth]{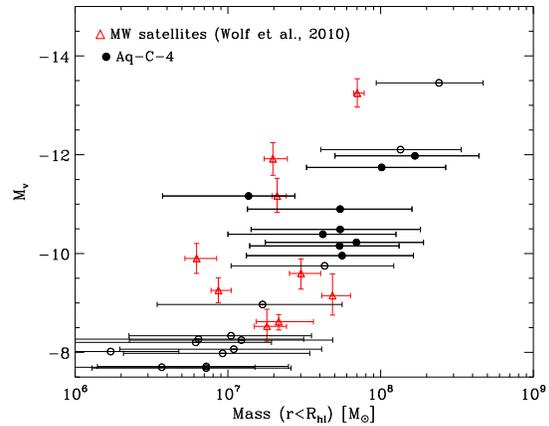}
\caption{The total mass contained within the projected half-light
  radius ($R_{hl}$) of simulated satellites, measured in the high
  resolution, dark matter only, Aquarius-C-2 simulation (open and
  filled black circles).  The radii are inferred from the observed luminosity-size relation
  (see text for details).  The filled circles indicate those
  satellites that are used in the statistical comparison with the
  observed data, illustrated in \fig{fig:KStest}.  The error bars on
  the simulated data points show the masses obtained by assuming
  half-light radii $1\sigma$ above and below the mean fitted values,
  where $\sigma$ is determined by the scatter about the fit to the
  observed data.  Red crosses with error bars are estimates from
  \citet{Wolf2010} for a selection of MW satellites.} 
\label{fig:masslum}
\end{figure}

The stellar kinematical properties of Local Group dwarf galaxies
provide an important test of the \lcdm cosmology.  Subhalos that form
in N-body simulations of MW-mass systems appear to have potentials
compatible with the stellar kinematics of the brightest MW satellites
\citep{Stoehr2002,StrigariFrenkWhite2010}.  Nonetheless, the analytic
calculations required to reach such conclusions necessarily include
simplifying assumptions.  Hydrodynamic simulations attempting to model
star formation self consistently in a cosmological setting are
inevitably some way behind the best N-body simulations in terms of
resolution and must also model uncertain baryonic physics on
sub-kiloparsec scales.  As such, our simulations are not suitable for
studying the detailed kinematics of the stars directly; instead, we
resort to a somewhat cruder comparison and ask whether our satellites
form in realistic potential wells, by comparing simulated and
observationally determined masses.

Historically there has been significant uncertainty associated with
determining satellite masses from observations.  Typically, estimates
are derived from the line-of-sight stellar velocity dispersion with three key
assumptions: i) the system is spherically symmetric, ii) stellar
orbits are isotropic and iii) the system is in equilibrium.  Two
recent studies have attempted a more general approach, with the aim of
reducing the systematic uncertainties.  Using an approach based on the spherical Jeans equation,
\citet{Walker2009} showed that for the brightest MW dSphs, the mass
within the projected half-light radius is robust to changes in the
anisotropy and underlying density profile.  This relation was
explained analytically by \citet{Wolf2010} who demonstrated that, if
the stellar velocity dispersion profile remains relatively flat in the
centre, as observations suggest \citep[e.g.,][]{Walker2007}, then the
uncertainty introduced by assuming a particular anisotropy is
minimised at the (3D) radius where the logarithmic slope of the
stellar number density profile, ${\rm -dln}n_{*}/{\rm dln}r = 3$.
They also showed that, for a range of realistic light profiles that
have been used to model the MW dSphs, this minimum lies close to the
(de-projected) half-light radius.  It is this radius, therefore, at
which we choose to compare the enclosed masses of satellites in the
simulations and observations.

In the previous subsection we described how aspects of our 
simulations, particularly the limitations of the subgrid treatment of
the ISM and the gravitational softening scale, can set an artificial
lower limit to the sizes of the stellar components of the satellites.
However, we also demonstrated that the luminosity function of the
 simulated satellites is close to that observed, the stellar mass in
each satellite is relatively well converged and stellar mass is found
to scale with subhalo mass similarly using an alternative modelling
technique.

With these checks in mind, we proceed with the assumption
that the cooling, star formation and feedback prescriptions in our
model result in a realistic stellar mass in each satellite, but that
stars form in a configuration that is too diffuse.  We then ask what
the projected half-light radius of each simulated satellite
\emph{should be} at a fixed luminosity, based on the observed sizes of the
brightest MW satellites.  We take a simple least-squares fit to the
data points in \fig{fig:sizes} (minimising the sum of the squared
differences in the magnitude coordinate) and compute the scatter in
the (log) radius coordinate about this line.  For each simulated
satellite, we assume a Gaussian distribution of possible sizes, with a
mean equal to the fit evaluated at the satellite's luminosity and
dispersion defined by the observed scatter.

Following this procedure we find that, for magnitudes
\magv$>-12$, our fit to the observed data implies sizes below the
scale at which softened gravitational forces become non-Newtonian in
the Aq-C-4 simulation, which leads to an underestimation of the
enclosed mass.  We choose instead to measure the mass of each
satellite in a much higher resolution dark matter only realisation of
the simulation, Aquarius-C-2, which has a smaller softening scale by a
factor of $\sim4$, such that the fitted half-light radii of satellites
down to \magv$\sim-7.5$ are larger than the force resolution.
The central masses measured in the higher resolution simulation are
typically forty to eighty percent higher for the ten brightest satellites, but the
difference can be a factor of three for satellites with \magv$\sim-8$.  Given the
results presented in \sect{sec:DMO}, we do not expect the omission of
baryons from Aquarius-C-2 to have had a large impact on the central
densities and hence the measured masses of these satellites.

\fig{fig:masslum} shows the mass enclosed within the mean half-light
radius chosen for each satellite, with error bars indicating the
masses corresponding to $\pm 1\sigma$ sizes.  Although the range of
plausible values is large, the brightest simulated satellites have
mean masses three to five times higher than the MW satellites of the same
luminosity.  While there is less of a discrepancy at fainter
magnitudes, our model seems to show a more gradual increase in 
luminosity with mass than is suggested by the observational data.  We
note that, starting from identical initial conditions to our Aq-C-4 simulation,
\citet{Wadepuhl2010} found that the mass-to-light ratios of their
satellites were typically higher than those quoted observationally by
a very similar factor and were also more discrepant in the most
massive satellites (see their Fig. 15).  In a hydrodynamic simulation
of the Local Group, again with resolution similar to our Aq-C-4,
\citet{Knebe2010b} found a similar result for satellites bound to
their MW and M31 analogues, with mass-to-light ratios a factor $\sim7$
too high.  We note, however, that they measured half-light radii for
their satellites using the star particles forming in their simulation,
which, as we have shown, can be too large when the scales associated
with star formation are not resolved.

To quantify the discrepancy in \fig{fig:masslum} statistically, we
construct multiple realisations of the half-light masses of the
simulated satellites by drawing sizes from the distributions described
above and computing the mass enclosed in the corresponding high resolution Aquarius-C-2
satellites.  We then combine the samples to define a model
distribution for the cumulative fraction of satellites with mass
larger than a given value and calculate the probability that the
observed masses could have been drawn from it, using a one-tailed
Kolmogorov-Smirnov (KS) test.

\begin{figure}
\includegraphics[width=0.4\textwidth]{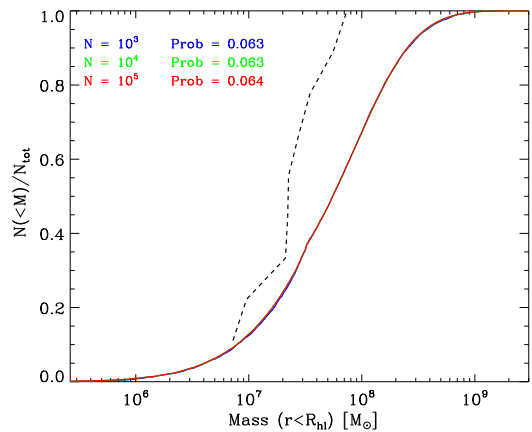}
\caption{The cumulative fraction of the $4{th}-12{th}$ brightest
  satellites as a function of the mass contained within the half-light
  radius in three model distributions (solid blue, green and red
  lines) and those derived for MW satellites by \citet{Wolf2010}
  (dashed line).  Labels in the top left indicate the number of times
  we repeat the process of drawing masses for our sample of nine
  satellites in order to define the model distribution, along with the
  probability that each model is consistent with the observed
  masses.}
\label{fig:KStest}
\end{figure}

In order to make a like-for-like comparison with the masses quoted by
\citet{Wolf2010}, we consider only the $4^{th}-12^{th}$ brightest
simulated satellites in our Aq-C-4 simulation that are within 280kpc,
corresponding to all of the classical MW satellites except the
Magellanic Clouds and Sagittarius, and including Canes Venatici (which
is approximately the same luminosity as Draco).  The sample is
indicated in \fig{fig:masslum} by filled circles.  Note that the third
brightest satellite in Aq-C-4 is in the midst of tidal disruption
(see \sect{sec:nef_sub}), a process that, as a result of small
differences in the orbits of the subhalos, is already complete in
Aquarius-C-2, hence no counterpart is found.  Although
\citet{Wolf2010} derived masses for 
fainter satellites, these correspond to simulated galaxies with fewer
than ten star particles, whose luminosities are uncertain in our
simulations and which we therefore choose to exclude.  In
\fig{fig:KStest} we plot cumulative model distributions, drawing
masses for the sample of nine satellites multiple times to define each
distribution, as indicated by the labels in the top left of the plot.
The probabilities that the observationally derived masses are
consistent with those distributions are found to be around six percent. 

As both \fig{fig:masslum} and \fig{fig:KStest} demonstrate, the masses
of the brightest simulated satellites are too high compared to those
derived for the MW satellites. This is another manifestation of the
problem recently highlighted by \citet{Boylan-Kolchin2011} who
compared the measured masses within the half-light radii of the same
satellites considered here with results from high-resolution
simulations of cold dark matter halos, including the Aquarius
suite. Assuming an NFW density profile \citep{Navarro1996,Navarro1997}, they
showed that the most massive subhalos in the simulations are too
concentrated to be able to host the brightest observed satellites.

The mismatch seen in \fig{fig:masslum} and in the results of
\citet{Boylan-Kolchin2011} could, in principle, be due to an underestimate of the
central masses of the observed satellites. However, for the errors
quoted by
\citet{Wolf2010} to be substantially underestimated would require
rather extreme variations in the anisotropy profile, which would be poorly
fit by their fairly general parameterised form.  This seems unlikely
to be the sole source of the disagreement between model and data.

While the discrepancy could be simply due to statistics, it might also
reflect a serious shortcoming either of the standard CDM cosmogony or
of current models of galaxy formation, such as those assumed in our
simulations.  A possible explanation of the discrepancy between the
mass-to-light ratios measured for the real and simulated satellites is
that the central dark matter densities predicted in the CDM model are
reduced by baryonic physics.  One mechanism for achieving this,
proposed by \citet{NavarroEkeFrenk1996}, is the condensation of a
dense baryonic component followed by the rapid expulsion of gas by
stellar feedback. The dark matter adjusts to this change in the
potential by developing a central ``core'', shifting the rotation
curve maximum to a larger radius and reducing the mass-to-light ratio
in the central parts.  This process does indeed appear to play an
important role in the evolution of one satellite in Aq-C-4 (see
\sect{sec:nef_sub}), which forms in the subhalo that has the largest
mass prior to accretion.  If this process is common, it is possible that it is not
seen here in less massive subhalos due to lack of resolution. 

A more radical explanation of the discrepancy is that the dark matter
consists of warm, rather than cold, particles. In this case, subhalos
of a given mass form later and have lower concentrations than in the
CDM model \cite[see][]{Navarro1997,HoganDalcanton2000}.  \citet{Lovell2011}
have recently shown explicitly that the masses and concentrations of
subhalos in a warm dark matter model agree well with the data.

\section{A star-dominated satellite} \label{sec:nef_sub}

\begin{figure*}
\includegraphics[width=0.92\textwidth]{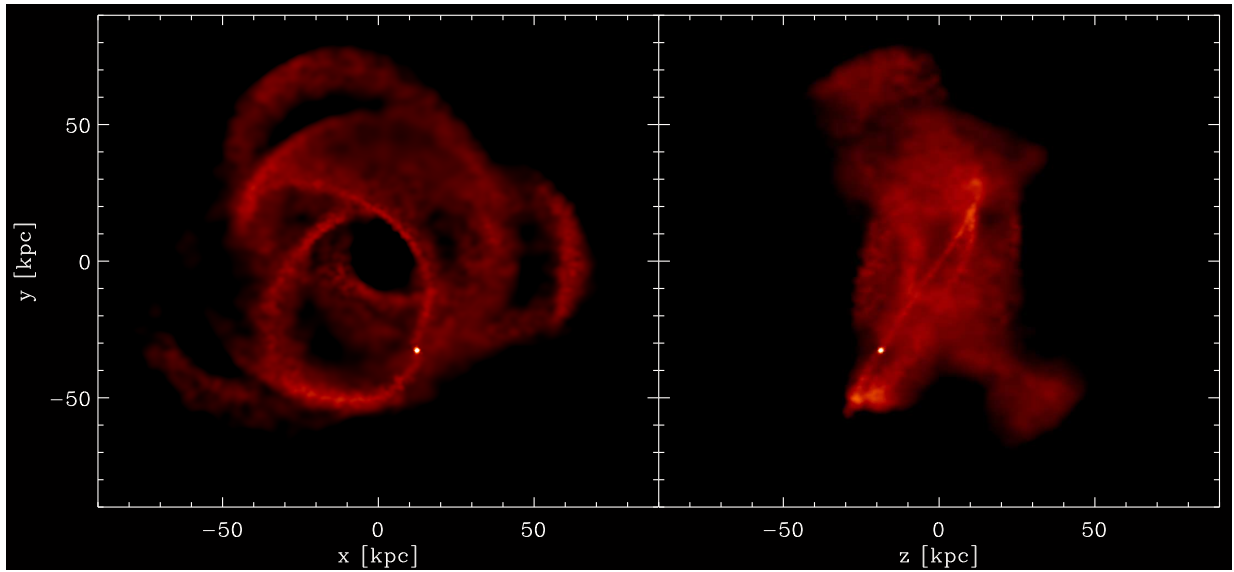}
\caption{Two orthogonal projections of the stellar stream associated
  with the satellite. The centre of the main halo is at the origin.  The
  stream is defined by selecting all stars associated with the
  satellite at the time of accretion and locating them at $z=0$. The
  surviving satellite is clearly seen as the bright concentrated
  object lying along the stream.}
\label{fig:sub33_stream}
\end{figure*}

The formation history of one of the satellite galaxies in our high
resolution hydrodynamical simulation is particularly interesting.  By $z=0$, we find that it
has become dominated by its stellar component, with a mass-to-light
ratio of $\sim2.4$ and its dark matter has become much less
concentrated than otherwise similar subhalos.  It appears as an outlier in
\fig{fig:sizes}, as it has a very small half-light radius for its
luminosity.  In this section, we briefly describe its formation
history and explain why it develops into such an unusual object.

At $z=0$ the satellite is, in fact, in the process of being tidally
disrupted and has a substantial stellar stream associated with it.
\fig{fig:sub33_stream} illustrates the structure of the stream in two
orthogonal projections centred on the main galaxy.  The dense stellar
nucleus of the satellite that remains identifiable as a bound structure is also
visible.  We track all star particles associated with the satellite at
the epoch when it is accreted and plot their projected mass density at
$z=0$.  The stripped stars account for the majority of the
stellar halo by mass. 

\begin{figure}
\includegraphics[width=0.4\textwidth]{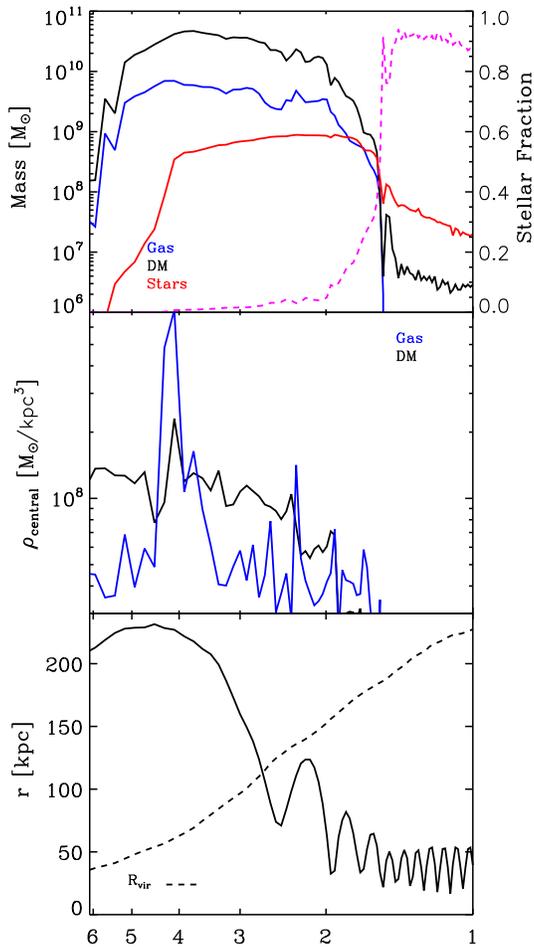}
\caption{Top panel: the mass in gas (blue), dark matter (black) and
  stars (red) gravitationally bound to the satellite's main
  progenitor, and the stellar fraction (magenta,dashed line, measured on the right
  vertical axis) as a function of redshift. Some of the small
  variations in the masses (and hence also the stellar fraction) at
  $z<1$ are due to the difficulty in identifying the subhalo's
  particles against the high background density at the centre of the
  main halo. Centre panel: the density of dark matter (black) and gas
  (blue) within the central 1kpc.  Bottom panel: the distance to the
  satellite from the centre of the main halo.  The dashed line
  indicates the virial radius of the main halo.}
\label{fig:sub33_fstar}
\end{figure}

The stream is a result of a fairly eccentric orbit with several close
pericentres, illustrated in the bottom panel of \fig{fig:sub33_fstar},
which shows the distance of the satellite from the centre of the main
galaxy as a function of redshift.  The dashed line indicates the
virial radius of the main halo. The accretion time is the point where
the two lines intersect.  The top panel tracks the mass in gas,
dark matter and stars bound to the satellite over the same redshift
interval, as well as the total mass fraction in stars.  At accretion,
it is the brightest satellite of the central galaxy, but only the
third brightest at $z=0$, as a result of the reduction in stellar mass
through tidal stripping.  The stellar fraction at accretion
($\sim0.02$) is fairly typical of the surviving satellites.  Note
that it is very common for the stellar fraction of a satellite to
increase with time after it is accreted, since the outer parts of the
dark matter halo are less tightly bound than the stars and hence more
susceptible to tidal stripping
\citep[e.g.][]{Penarrubia2008b,Sawala2011}.  The middle panel of
\fig{fig:sub33_fstar} shows the evolution of the central ($\rm r < 1kpc$) density of
the subhalo in gas and dark matter, both of which drop sharply when the satellite is
close to pericentre.  A decline is also evident after $z\sim3$, well before the satellite
is accreted, the origin of which we discuss in more detail below.

During the first few orbits, the stellar component remains unaffected
while the dark matter lying beyond the radial extent of the stellar
component is stripped.  In fact, some of the dark matter particles
with pericentres \emph{within} the stellar component are also stripped,
as a result of the dark matter having a higher radial velocity
dispersion than the stars.  The final masses of the stellar and dark
matter components are factors of $\sim50$ and $2\times10^4$ lower than
their peak values respectively.

\begin{figure}
\includegraphics[width=0.4\textwidth]{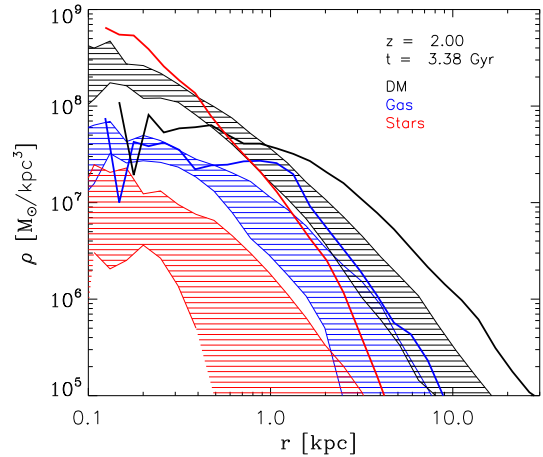}
\caption{The dark matter (black), stellar (red) and gas
  (blue) density profiles of the galaxy at $\rm z\sim2$, when it is
  first accreted as a satellite.  Regions of the same colour indicate
  the spread of values ($\pm1\sigma$) in each radial bin for the nine most
  massive surviving satellites at $z=0$.}
\label{fig:sub33_dens}
\end{figure}

The extent to which the two components are stripped is strongly
affected by their radial density profiles, which are shown in
\fig{fig:sub33_dens} at the time of accretion.  The overplotted
regions indicate the range of densities ($\pm1\sigma$) in each bin for
the nine most massive surviving satellites.  Clearly, the stellar
component of this galaxy is unsually concentrated relative to those
other galaxies, whilst the dark matter and gas have shallower than
average central density profiles.  It is unclear how much effect the
gravitational softening has in this respect, since forces begin to
become sub-Newtonian on scales less than $\rm\lsim720pc$, but we note
that the profiles also differ outside that radius.  The highly
`cusped' stellar profile allows the central stellar nucleus to resist
the strong tidal forces that unbind the majority of the dark matter.
It also accounts for the unusually small half-light radius shown in
\fig{fig:sizes}.

The origin of these density profiles is related to the satellite's
violent formation history.  In a series of major mergers at $z\sim3$,
gas is funnelled to the centre of the main progenitor, initiating an
intense burst of star formation that gives rise to a highly
concentrated stellar distribution.  The subsequent burst of feedback
energy rapidly removes a large fraction of the gas and leads to a fall
in the mean binding energy of the central dark matter.  This episode
is clearly visible in the bottom panel of \fig{fig:sub33_fstar}, which
shows the central gas density drop as it is expelled by feedback and
turned into stars, followed by a decline in the dark matter density in response
to the change in the potential.  This sequence of events is effectively the process
originally proposed by \citet{NavarroEkeFrenk1996}.  The reduced
binding energy of its central dark matter, along with an extreme
orbit, combine to produce the unusual properties of this satellite at $z=0$. 

%%%%%%%%%%%%%%%%%%%%%%%%%%%%%%%%%%%%%%%%%%%%%%%%%%%%%%%%%%%%%%%%%%%%%%%%%%%%%%%%
%%%%%%%%%%%%%%%%%%%%%%%%%%%%%%%%%%%%%%%%%%%%%%%%%%%%%%%%%%%%%%%%%%%%%%%%%%%%%%%%
\section{Conclusions} \label{sec:CONCLUSIONS}

We have investigated the formation and evolution of a Milky Way-like
satellite system in an SPH simulation over three levels of resolution,
in which particle masses vary by a factor of 64.  The properties of
our simulated satellites show relatively good numerical convergence,
with the final stellar masses typically agreeing to within a factor of
two, and always to within a factor of six.  We also compared to an
independent estimate of the stellar mass expected to form in each
subhalo, using the semi-analytic model of \citet{Cooper2010}.  The two
theoretical techniques produce a similar ranking of the subhalos by
stellar mass, although our simulations typically form a higher mass of
stars by a factor of between two and six.  This discrepancy may be partly
explained by the assumption in the semi-analytic model that gas is
stripped instantaneously when a galaxy becomes a satellite.  The mass
evolution in gas, stars and dark matter of each satellite agrees well
between resolutions, except that gas is stripped more rapidly at lower
resolution following accretion onto the main halo.  Poorer force
resolution causes gas particles to be more loosely bound to the
subhalo and hence more susceptible to ram pressure stripping.  This
phenomenon may account for many of the differences in the final
stellar masses between resolutions.

By comparing the dark matter halos of our satellite galaxies to those
that form in a dissipationless version of the same simulation, we were
able to quantify the expected impact of baryons on the phase-space
structure of the dark matter.  Due to small deviations in satellite
orbits between different realisations of the same halo, it is
necessary to make this comparison when the satellite first falls in
rather than at $z=0$.  Although in some radial bins, in a few
subhalos, the density and velocity dispersion profiles are found to
change by $\sim30$ percent, the differences were typically less than
10~percent.  With the caveat that the resolution of our simulation may
limit the magnitude of such effects, we conclude that baryons have a
relatively small impact on the structure of the dark matter halos of
satellite galaxies around MW-like hosts.

Our model provides a reasonable match to the faint end of the Local
Group satellite luminosity function averaged between the Milky Way and M31,
although there is a slight deficit at the bright end, with no LMC
analogue.  However, SDSS data \citep{Liu2010,Guo2011}
suggest that it is quite common for galaxies with luminosities like
the MW to have no satellites as bright as the LMC and SMC.

Due to the limitations of the spatial resolution and the implementation of
baryonic physics in our simulations, particularly the modelling of
the multiphase ISM, stars do not form in sufficiently concentrated
distributions to match the half-light radii of Local Group satellites.
However, the reasonable agreement between the stellar masses in simulations with
different resolution and different modelling techniques, combined with
the match to the observed satellite luminosity function, suggest that
the baryonic mass that is able to cool in each (sub)halo and form
stars is realistic. 

In order to test whether satellites of a given luminosity form in
halos with masses consistent with those observed, we compare their
`half-light masses' with values derived for a selection of the
brightest MW satellites by \citet{Wolf2010}.  For this comparison,
each satellite is assigned a distribution of half-light radii from the
best fit to the observed luminosity-size relation and its variance.
In the hydrodynamical simulation, the gravitational softening is
comparable to these fitted half-light radii, so we instead measure the
masses in a much higher resolution, dark matter only realisation of
the simulation.  We hence explicitly ignore any effects baryons may
have had on the central density profiles, which, in any case, the
results in \sect{sec:DMO} suggest are small.  The large scatter about the observed
relation translates into a broad range of possible masses for each
simulated satellite, but nonetheless, the mean masses for the
brightest examples ($Mv<-11$) are about three to five times higher
than their observed counterparts.  The observed mass-luminosity
relation seems to be somewhat steeper than that produced by our model,
although due to the small sample sizes, the slope of the relation and
the scatter about it are relatively poorly defined in both cases.  A
KS test, taking into account the uncertainties in the half-light radii
assigned to the simulated satellites, returns a six percent
probability that the observed masses could have been drawn from the
distribution defined by the simulation data.

Although the apparent disagreement between the simulations and the
data could be simply due to statistics, there are also a number of
plausible physical explanations.  It could be that baryonic processes
significantly reduce the central dark matter densities of satellite
galaxies. Possible mechanisms to achieve this include, for instance, a
sudden change in the local potential, induced by the rapid expulsion
of baryonic mass  through stellar feedback
\citep[e.g.][]{NavarroEkeFrenk1996} or heating due to bulk motions of
dense clumps of gas \citep[e.g.][]{Mashchenko2006}.  The results in
\sect{sec:DMO} imply that, if such processes are important, they are
either not resolved in the Aq-C-4 simulation, or are not properly captured
by our feedback prescription, except in one case, which happens to be
the most massive subhalo at accretion.  Less concentrated dark
matter profiles would also result if the dark matter consists of warm,
rather than cold particles \citep[e.g.][]{Lovell2011}.

The broad range of possible explanations of the discrepancy
highlighted by our results illustrates how uncertain our understanding
of galaxy formation still is on the scale of dwarf
galaxies. Determining which, if any, is correct will be of critical
importance in assessing the viability of the CDM cosmology and the
success of galaxy formation models.

%%%%%%%%%%%%%%%%%%%%%%%%%%%%%%%%%%%%%%%%%%%%%%%%%%%%%%%%%%%%%%%%%%%%%%%%%%%%%%%%
%%%%%%%%%%%%%%%%%%%%%%%%%%%%%%%%%%%%%%%%%%%%%%%%%%%%%%%%%%%%%%%%%%%%%%%%%%%%%%%%
\section*{Acknowledgments}
We thank Joop Schaye for helpful comments in the early stages of this work and
Adrian Jenkins for creating the initial conditions for the
simulations.  We are also grateful to Andrew Cooper for giving us
access to his semi-analytic satellite data and for helpful
discussions. OHP acknowledges the receipt of an STFC studentship. TO
acknowledges financial support by Grant-in-Aid for Scientific Research
(S) by JSPS (20224002) and by Grant-in-Aid for Young Scientists
(start-up: 21840015).  Simulations associated with this work were run
on the IBM pSeries Power6 at the Rechenzentrum, Garching, the
Cosmology Machine at the Institute for Computational Cosmology (ICC)
in Durham, the Cray XT4 at the National Astronomical Observatory of
Japan's Centre for Computational Astrophysics and at the Centre for
Computational Sciences in the University of Tsukuba.  CSF acknowledges
a Royal Society Wolfson research merit award.  We thank the DEISA
Consortium (www.deisa.eu), co-funded through the EU FP6 project
RI-031513 and the FP7 project RI-222919, for support within the DEISA
Extreme Computing Initiative.  This work was supported in part by a
STFC rolling grant to the ICC and ERC Advanced Investigator grant
267291 COSMIWAY.

\bibliographystyle{mn2e}
\bibliography{bibliography}

\label{lastpage}
\end{document}